\def\ir{{\rm I}\hskip-.2em{\rm R}}

\def\ra{\rightarrow}
\def\half{\textstyle{\frac{1}{2}}}
\def\d{\partial}
\def\o{\overline}
\def\b{\begin{eqnarray*}}     %takes no eqn numbers
\def\e{\end{eqnarray*}}       %takes no eqn numbers
\def\bn{\begin{eqnarray}}     %takes eqn numbers
\def\en{\end{eqnarray}}       %takes eqn numbers
\def\<{\langle}
\def\>{\rangle}
\def\dn{d^{n}\!x}
\def\ds{d^{s}\!x}
\def\dny{d^{n}\!y}
\def\dsy{d^{s}\!y}
\documentstyle[12pt]{article}
\title{Poisson Distributions for Sharp-Time \\Fields: Antidote for
Triviality}
\author{John R. Klauder\\
Departments of Physics and Mathematics\\
University of Florida\\
Gainesville, Fl  32611}
\begin{document}
\maketitle
\begin{abstract}
Standard lattice-space formulations of quartic self-coupled Euclid-ean
scalar quantum fields become trivial in the continuum limit for sufficiently
high space-time dimensions, and in particular the moment generating
functional for space-time smeared fields becomes a Gaussian appropriate to
that of a (possibly generalized) free field.  For sharp-time fields this
fact implies that the ground-state expectation functional also becomes
Gaussian in the continuum limit. To overcome these consequences of the
central limit theorem, an auxiliary, nonclassical potential is appended to
the original lattice form of the model and parameters are tuned so that a
generalized Poisson field distribution emerges in the continuum limit for
the ground-state probability distribution. As a consequence, the sharp-time
expectation functional is infinitely divisible, but the Hamiltonian operator
is such, in the general case, that the generating functional for the
space-time smeared field is not infinitely divisible in Minkowski space.
This feature permits the models in question to escape a manifestly trivial
scattering matrix imposed on all infinitely divisible covariant Minkowski
fields. Two sequentially related proposals for an alternative lattice
formulation of interacting covariant models in four and more space-time
dimensions are analyzed in some detail.
\end{abstract}
\section{Introduction and Overview}
One of the more useful techniques to generate a Euclidean scalar quantum
field theory is to take the continuum and infinite volume limit of a model
theory defined on a finite space-time hypercubic lattice. Let the expression
  $$\<e^{\Sigma h_k\phi_k a^n }\>\equiv \int e^{[\Sigma h_k\phi_k a^n -
{\cal W}(\phi)]}\,\Pi d\phi_k$$
schematically represent the generating functional for such a lattice theory.
Here $\{h_k\}$ is an external field, $k=(k_1,k_2,\ldots,k_n),\;k_j\in {\bf
Z}\equiv\{0,\pm 1,\pm 2,\dots\}$, labels a lattice site, $a$ denotes the
lattice spacing, and $n$ denotes the space-time dimension. The expression
${\cal W}(\phi)$ denotes the lattice action, which for a $\varphi^4_n$
model, for example, is conventionally taken to be
  $${\cal W}(\phi)=\half Y(a)\Sigma (\phi_{k^*}-\phi_k)^2a^{n-2}+\half
m_0^2(a)\Sigma\phi_k^2a^n+g_0(a)\Sigma\phi_k^4a^n,$$
where $k^*$ denotes one of half of the nearest neighbors to $k$ and the
summations run over the sites and nearest neighbors as needed on a finite
size lattice. The continuum limit includes an increase in the number of
lattice sites as well as sending $a\ra0$. To facilitate convergence to
potentially interesting results, cutoff-dependent coefficients ($Y,\,m_0,$
and $g_0$) have been included. For $n=2,\,3$ the resultant theory that
emerges in the continuum limit is nontrivial (non-Gaussian) and is in accord
with that computed from a (renormalized) perturbation theory in the coupling
constant $g_0$ \cite{glj}. When $n\geq 5$, and possibly for $n=4$ as well,
the theory that emerges in the continuum limit is trivial (Gaussian), to wit
  $$\<e^{\int h(x)\phi(x)\,\dn}\>=e^{\,\frac{1}{2}\!\int h(x)C(x-y)h(y)\,
\dn\,\dny}\;,$$
which corresponds to a Minkowski theory that is a (possibly generalized)
free field \cite{gal}. The manifest non-Gaussian character of the model on
the lattice is subsumed into a possible influence on the correlation
function $C$ in the Gaussian limiting behavior.

The limiting behavior described above for $n\geq5$, and possibly $n=4$ as
well, is characteristic of the general behavior that goes under the name of
the Central Limit Theorem (CLT)\cite{clt}. Although the simplest
illustrative examples of a CLT behavior are for independent identically
distributed random variables, neither independence nor identity of the
various distributions are essential for its validity. If we accept for the
further sake of argument that a CLT mechanism is behind the Gaussian
limiting behavior for $n\geq4$ (including $n=4$ for convenience), then we
are also drawn to consider Poisson distributions as they are the very kind
that resist the vise grip of the CLT and lead to non-Gaussian and thereby
potentially nontrivial continuum limits. By Poisson distributions we mean
all those that go under the name Poisson, compound Poisson, or generalized
Poisson distributions \cite{def}, which are described below as needed. It is
the latter of these distributions that will be of physical interest in the
present work.

Once the possibility of Poisson distributions is raised, there are (at
least) two ways to proceed. From the point of view of Euclidean quantum
field theory---where we presently find ourselves---a natural procedure,
which we shall initially discuss, suggests itself, but we shall find it
wanting in that although the Euclidean theory may be non-Gaussian the
associated Minkowski theory is forced to have trivial scattering \cite{buc}.
Our second construction of a Poisson distribution is qualitatively different
than the first one, and nontrivial scattering in the Minkowski theory is not
manifestly excluded. Although, in hindsight, the second approach may appear
more natural, both standpoints are included for completeness. We now outline
the two alternative approaches in more detail.

One may first be tempted to insist that the entire space-time dependent
field distribution should be made into a generalized Poisson distribution in
Euclidean space.  This change of distribution is presumed to arise by the
addition of an auxiliary, nonclasssical $[O(\hbar^2)]$ potential ${\o P}$ to
the conventional lattice action. Assuming for the moment we could find such
a ${\o P}$ let us examine its consequences for potential nontriviality. In
particular, we suppose that both the Euclidean and Minkowski fields have the
property of {\it infinite divisibility} \cite{luk}\cite{heg}. For the
Minkowski field this means that the generating functional for the smeared
field operator
  $$\varphi(h)\equiv\int h(x)\varphi(x)\,\dn$$
has a form implicitly given by
  $$\<0|e^{i\varphi(h)}|0\>\equiv\lim_{R\ra\infty}
\exp\{\,R\<0_R|\,[e^{i\varphi_R(h)}-1]|0_R\>\},$$
where $|0_R\>$ is a normalized state and $\varphi_R(h)$ is a field operator
for each $R$. As a consequence of this structure it follows, for example,
that
  $$\<0|\varphi(f_1)\varphi(f_2)\varphi(f_2)\varphi(f_1)|0\>^T\geq 0\;.$$
Combined with Schwarz's inequality we have
\b
0&\leq&|\<0|\varphi(g_1)\varphi(g_2)\varphi(f_2)\varphi(f_1)|0\>^T|^2\\
&\leq&\<0|\varphi(g_1)\varphi(g_2)\varphi(g_2)\varphi(g_1)|0\>^T\\
&&\times\<0|\varphi(f_1)\varphi(f_2)\varphi(f_2)\varphi(f_1)|0\>^T \;,
\e
which for the asymptotic states of the Haag-Ruelle scattering theory implies
that
\b
0&\leq&|\<0|\varphi_{out}(g_1)\varphi_{out}(g_2)\varphi_{in}(f_2)
\varphi_{in}(f_1)|0\>^T|^2\\
&\leq&\<0|\varphi_{out}(g_1)\varphi_{out}(g_2)\varphi_{out}(g_2)
\varphi_{out}(g_1)|0\>^T\\
&&\times\<0|\varphi_{in}(f_1)\varphi_{in}(f_2)\varphi_{in}(f_2)
\varphi_{in}(f_1)|0\>^T =0\;.
\e
The vanishing of this expression implies an absence of two-particle elastic
scattering. Similar arguments may be extended to multi-particle scattering
including particle production and annihilation \cite{buc}. Accepting the
hypothesis  that all asymptotic states are covered in this way leads to a
unit $S$-matrix. Thus, insisting that the entire space-time dependent field
distribution be a generalized Poisson distribution leads to trivial
scattering, and so this particular hypothesis loses its appeal. However,
there is also another route to follow.

Let us assume for the sake of argument that the sharp-time field is well
defined (a discussion of lifting this assumption appears in the next
Section). We observe that if the space-time smeared field has a Gaussian
generating functional then this property is inherited by the sharp-time
field as well. If we set
  $$h(x)=g({\bf x})\delta (t)\;,\;\;\;\;{\bf x}\in\ir^s\;,\;\;\;\;s\equiv
n-1\;,$$
then, for $n\geq4$,
  $$\<e^{\int g({\bf x})\phi({\bf x})\,\ds}\>=e^{\,\frac{1}{2}\!
\int g({\bf x})C({\bf x}-{\bf y})g({\bf y})\,\ds\,\dsy}\;.$$
At sharp time, e.g., $t=0$, the Euclidean theory expectation coincides with
the Minkowski theory expectation, and thus it follows that
  $$\<0|e^{i\varphi(g)}|0\>=e^{-\frac{1}{2}\!\int g({\bf x})C({\bf x}-
{\bf y})g({\bf y})\,\ds\,\dsy}\;,$$
where we have changed $g$ to $ig$ for later convenience. We may diagonalize
the field at $t=0$ so that
  $$\<0|e^{i\varphi(g)}|0\>=\int
e^{i\phi(g)}|\Psi_G\{\phi\}|^2\,\delta\phi$$
where we have adopted a formal notation for a straightforward Gaussian
functional integral. The expression $\Psi_G\{\phi\}$ represents the ground
state functional, a Gaussian eigenstate of the Hamiltonian operator.

The Gaussian character of the continuum ground state contrasts with the {\it
non}-Gaussian ground state of the lattice Hamiltonian operator prior to
taking the continuum limit. The lattice Hamiltonian may be readily deduced
from the lattice action, and it has the form
\b {\cal H}_a  &=&-{\half} Y^{-1}(a)
a^{-s}\Sigma'\frac{\d^2}{\d\phi_k^2}+{\half}
Y(a)\Sigma'(\phi_{k^*}-\phi_k)^2a^{s-2}\\
&&+{\half} m_0^2(a)\Sigma'\phi_k^2a^s+g_0(a)\Sigma'\phi_k^4a^s+
{\rm const.}\;,  \e
where we assume that $g_0(a)>0$ and that the prime on various sums indicates
that only lattice sites with the same ``time'' value (equal $k_1$, say) are
included. The ``const.'' is adjusted so that the ground state $\Psi_a(\phi)$
of the lattice Hamiltonian satisfies
  $${\cal H}_a\Psi_a(\phi)=0\;.$$
In view of the nonvanishing quartic interaction, $\Psi_a(\phi)$ can not be a
Gaussian, and as a consequence the characteristic functional
  $$\int e^{i\Sigma'\phi_kg_ka^s}|\Psi_a(\phi)|^2\,\Pi'\,d\phi_k$$
is {\it non}-Gaussian although, as noted above, the continuum limit of this
lattice characteristic functional is Gaussian. Thus, we may reasonably
conclude that the distribution of field values at sharp time (e.g., $t=0$)
succumbs to the CLT, and in the continuum limit passes from a non-Gaussian
to a Gaussian distribution.

Our strategy at this point is to add an (alternative) auxiliary,
nonclassical $[O(\hbar^2)]$ potential $P$ to the lattice Hamiltonian which
has the effect of changing the ground state in such a way that in the
continuum limit the characteristic functional for the sharp-time field
corresponds to a generalized Poisson distribution. In symbols, in the first
approach we adopt, we modify the lattice Hamiltonian so that its
ground-state characteristic functional has the form given by
$$  \int e^{i\Sigma'\phi_kg_ka^s}|\Psi_a(\phi)|^2\,\Pi'd\phi_k
 =\exp\{-\int[1-\cos(\Sigma'\phi_kg_ka^s)]
|\psi_a(\phi)|^2\,\Pi'd\phi_k\}\;,  $$
where
  $$\int|\psi_a(\phi)|^2\,\Pi'd\phi_k=\infty\;,$$
and insist, moreover, in the continuum limit and written formally, that
 \b
  \<0|e^{i\varphi(g)}|0\>&=&\int
 e^{i\phi(g)}|\Psi_P\{\phi\}|^2\,\delta\phi\\
  &=&\exp(-\int\{1-\cos[\phi(g)]\}|\psi\{\phi\}|^2\,\delta\phi)\;,
\e  where
  $$\int|\psi\{\phi\}|^2\,\delta\phi=\infty\;.$$
In writing these expressions we have made explicit use of the expected
symmetry $\psi_a(-\phi)=\psi_a(\phi)$ appropriate to the ground state in
order to introduce the ``cosine''. The expression in the exponent in both of
the last equations is often termed the {\it second characteristic}
\cite{luk}; it also serves as the generator of the truncated equal-time
correlation functions, when such moments exist. We suppose that the
characteristic functional---and thus the second characteristic---is well
defined for a wide class of test functions $g$, say $g\in {\cal S}(\ir^s)$,
the Schwartz space of $C^\infty$ functions of rapid decrease. However, as
already indicated, we will insist that
$\int|\psi_a(\phi)|^2\,\Pi'd\phi_k=\infty$ as well as
  $\int|\psi\{\phi\}|^2\,\delta\phi=\infty$,
conditions which characterize a {\it generalized} Poisson distribution for
the lattice and in the continuum limit, respectively, as this divergence is
necessary and sufficient in the present case for the smeared sharp-time
field operator $\varphi(g)$, on the lattice and in the continuum, to have a
purely absolutely continuous spectrum. The contrary situation for which
$\int|\psi_a(\phi)|^2\,\Pi'd\phi_k<\infty$, or especially
$\int|\psi\{\phi\}|^2\,\delta\phi<\infty$, conditions which describe a
compound Poisson distribution, will be unsuitable.

Once having ensured that the sharp-time field has a generalized Poisson (and
non-Gaussian) distribution, we choose the associated functional
$\Psi_P\{\phi\}$ to be the ground state of the continuum Hamiltonian and,
based on that choice, attempt to build multi-time correlation functions in a
conventional manner. Insofar as the sharp-time correlation functions are
non-Gaussian, one may reasonably expect the multi-time correlation functions
to be non-Gaussian as well. Since the present construction invokes an
auxiliary interaction to generate a Poisson distribution only for the
sharp-time fields, there is no reason, in general, to expect that the
relativistic field will have a Poisson distribution for all space {\it and}
time thereby avoiding manifest triviality of the $S$-matrix as discussed
above.

Faced with the difficulty of analytically determining $\Psi_a(\phi)$ to
carry out this program we introduce a second model designed so that the
continuum limit of the time zero field, at least for several test cases, is
identical to that based on the first model. In the second model we
approximate $\Psi_a(\phi)$ directly thereby enabling us to recover the
auxiliary nonclassical potential $P$ directly. Given the auxiliary potential
we propose a specific form of a lattice action that we believe embodies our
proposals for an alternative and nontrivial quantization of self-coupled
scalar fields such as $\varphi^4_n$ for $n\geq4$ (and potentially an
alternative quantization for $n=2$ or $3$ as well).
\subsection{Illustrative Toy Model}
It is instructive at this point to introduce a ``toy'' model to more easily
visualize some aspects of the previous discussion. Let us consider the
well-known Cauchy distribution in one dimension given by its characteristic
function
  \b C(p)\equiv e^{-b|p|}&=&\int \frac{b\, e^{ipx}}{\pi(x^2+b^2)}\,dx\\
    &=&\exp\{-b\int[1-\cos(px)]\frac{\,dx}{\pi x^2}\}\;.  \e
Here $b>0$ denotes an arbitrary scale parameter. The spectrum of $x$ is
absolutely continuous as signified by the probability density
$b/[\pi(x^2+b^2)]$. Observe the nonintegrable singularity in the second
characteristic weight function, i.e.,  $\int\,dx/(\pi x^2)=\infty$. The two
weight functions involved are, of course, related to one another. In
particular,
  \b |p|=\lim_{b\ra0}\,b^{-1}(1-e^{-b|p|})&=&\lim_{b\ra0}\int(1-e^{ipx})
\frac{\,dx}{\pi(x^2+b^2)}\\
  &=&\int[1-\cos(px)]\frac{\,dx}{\pi x^2}\;.   \e

While analytically simple, this toy model is special because its moments do
not exist. A closely related family of toy models is determined by the
characteristic function
  \b
  C_\rho(p)=e^{-b\sqrt{p^2+\rho^2}+b\rho}&=&\int
e^{ipx}\frac{b\rho}{\pi}\frac{K_1(\rho\sqrt{x^2+b^2}\,)}{\sqrt{x^2+b^2}}
e^{b\rho}\,dx\\
   &=&\exp\{-b\int[1-\cos(px)]\frac{\rho K_1(\rho|x|)}{\pi|x|}\,dx\}\;,
\e
where $\rho>0$ is another parameter at our disposal \cite{cam}. The
conventional modified Bessel function $K_1(z)\simeq \sqrt{\pi/2z}\,e^{-z}$
for $z\gg1$, ensuring that all moments of the distribution exist. Moreover,
since $K_1(z)\simeq 1/z$ for $0<z\ll1$, it follows that as $\rho\ra0$ these
distributions converge to the Cauchy case. For any $\rho>0$ observe that for
{\it small} $x$ the second characteristic weight function is effectively
identical, i.e., $b/(\pi x^2)$, to that of the Cauchy distribution. Thus we
are led to the notion of a set of different distributions all of which have
the same singularity structure of the second characteristic weight function
at small arguments but which differ from each other by their behavior for
intermediate and large arguments. In the present case such a set includes,
e.g., all distributions of the form
  $$\exp\{-b\int[1-\cos(px)]\frac{F(x)}{\pi x^2}\,dx\}\equiv\int
e^{ipx}G(x)\,dx\;,$$
where $F\in C^2,\;F(0)=1,\;F\geq0,$ and $\int[F(x)/(1+x^2)]\,dx<\infty$.
Although we can not analytically specify the $L^1$ nonnegative weight
function $G$ of the characteristic function in the general case, we know
that it exists. The characterization of a distribution by means of its
second characteristic will be an important tool in our study.

One aspect of our approach that does not have an analog in this toy model is
the relation of the second model to the first, namely the fact that these
two lattice field theories have different lattice starting points but equal
continuum limits.

Before closing the discussion of our toy models it is worthwhile
illustrating our suggested procedure for constructing multi-time correlation
functions. For simplicity we confine ourselves to the Cauchy model (even
though an entirely analogous procedure could be carried through for the toy
model with finite moments). In particular, if we identify
  $$\<0|e^{ipQ}|0\>\equiv e^{-b|p|}=\int e^{ipx}\frac{b\,dx}{\pi(x^2+b^2)}$$
as the ground-state expectation function, then we are led to represent the
Hamiltonian operator $\cal H$ by
  \b {\cal H}_b&\equiv&-{\half}\frac{\d^2}
{\d x^2}+{\half}(x^2+b^2)^{1/2}\frac{\d^2}{\d x^2}(x^2+b^2)^{-1/2}\\
    &=&-{\half}\frac{\d^2}{\d x^2}+\frac{2x^2-b^2}{2(x^2+b^2)^2}\;.   \e
For any $b>0$ it is seen that this Hamiltonian has only one normalizable
eigenstate, the ground state with zero energy; this paucity of normalizable
eigenstates is related to the absence of moments for this model. It follows
that
 $$\<0|e^{ip'Q}
e^{-i{\cal  H}t}e^{ipQ}|0\>\equiv\frac{b}{\pi}\int\frac{e^{ip'x}}
{\sqrt{x^2+b^2}}e^{-i{\cal H}_bt}\frac{e^{ipx}}{\sqrt{x^2+b^2}}\,dx\;,$$
with analogous expressions holding for higher-order multi-time correlations.
The procedure illustrated here is analogous to that to be followed in our
discussion of $\,\varphi^4$ models in Sections 2 and 3.

On the other hand, for the sake of completeness, we should also illustrate
an alternative dynamical extension analogous to the first proposal to have a
Poisson distribution for fields over all space and time. In this alternative
dynamical extension we modify the second characteristic so that
 $$\<0|e^{ip'Q}e^{-i{\cal H}t}e^{ipQ}|0\>
\equiv\exp[\,\frac{b}{\pi}\int\frac{1}{|x|}
(e^{ip'x}e^{-i{\sf h}t}e^{ipx}-1)\frac{1}{|x|}\,dx]\;,$$
where in the present case
  \b {\sf h}&\equiv&-{\half}\frac{\d^2}{\d x^2}+{\half}|x|
\frac{\d^2}{\d x^2}|x|^{-1}\\
  &=&-{\half}\frac{\d^2}{\d x^2}+\frac{1}{x^2}\;.  \e
At $t=0$, of course, the two dynamical extensions coincide. The Hamiltonian
{\sf h} has a positive, purely absolutely continuous spectrum, and as a
consequence, the asymptotic behavior of both dynamical expressions exhibits
clustering as $t\ra\pm\infty$. Thus on general grounds both dynamical
expressions may seem viable. However, as argued earlier, one of these
extensions leads to manifestly trivial scattering in the field theory case.
Consequently, our further discussion will be aimed at converting the
sharp-time field distribution into a generalized Poisson distribution and
proceeding in analogy with the former of the two dynamical toy models.

\subsection{Summary of Proposed Lattice Model for $\varphi^4_n$}
Here we summarize the results of the next two sections and present in as
concrete a form as possible our proposal for a lattice model designed to
lead to a nontrivial continuum limit and to have the possibility of
nontrivial scattering as a relativisitic theory. In the language with which
the conventional lattice models are formulated, our proposal is very much
the same save for two aspects. The first difference is the presence of the
auxiliary, nonclassical potential that enforces a generalized Poisson
distribution on the sharp-time field distribution, and the second difference
is the wholly new form taken by the bare parameters of the theory in terms
of the renormalized (or finite, nonzero) parameters that eventually
determine the theory properly.

The model is characterized by the Euclidean lattice form of the correlation
generating function as given by
 \b
  S(h)&=&N_0\int\exp[\Sigma\, h_k\phi_ka^n-\half
Y(a)\Sigma(\phi_{k^*}-\phi_k)^2a^{(n-2)}-\half m_0^2(a)\Sigma\phi_k^2a^n\\
& &\;\;\;\;\;\;\;\;\;\;-g_0(a)\Sigma\phi_k^4a^n-\Sigma\,
P(\phi_k,a)a^n]\:\Pi d\phi_k\;.  \e
In this expression \cite{k94}
  $$P(\phi_k,a)=\frac{{\sf A}(a)\phi_k^2-{\sf B}(a)}
{[\phi_k^2+{\sf C}(a)]^2}\;,$$
where
 \b
    &&{\sf A}(a)= Y(a)^{-1}a^{-2s}\,8^{-1}(1+\xi)(3+\xi)     \;,\\
    &&{\sf B}(a)= Y(a)^{-1}a^{-2s}\,4^{-1}(1+\xi){\sf C}(a)      \;,\\
     &&{\sf C}(a)= M^{n-2}(Ma)^{2s(1-\xi)/\xi}     \;.    \e
In these relations $\xi>0$ is a free parameter labeling inequivalent
quantizations; as we shall show, the cases $0<\xi<2$ and $2\leq\xi$ are
qualitatively different. For the present, as well as for the most part, we
shall confine attention to $0<\xi<1$. Our analysis indicates that the bare
parameters are related to renormalized parameters according to the relation
\b  m_0^2(a)\!\!\!\!\!\!\!&&=(Ma)^{n-2}\,m^2\;,\\
    g_0(a)\!\!\!\!\!\!\!&&=(Ma)^{3n-4}g\;,\\
    Y(a)\!\!\!\!\!\!\!&&=(Ma)^n\;,   \e
where $M$ denotes a mass parameter chosen to ensure that engineering
dimensions are preserved. Observe that these three bare parameter
renormalizations are {\it multiplicative} as opposed to {\it subtractive;}
this consequence is a direct reflection of the profound difference that the
auxiliary potential introduces into the integrand.  Moreover, all three bare
parameters are {\it small} when $Ma\ll1$, i.e., as the continuum limit is
approached. This means that in the integration variables indicated the
essential support of $\phi$ extends roughly to $O((Ma)^{-s})$, namely to
large values. On the other hand, the contribution of the auxiliary potential
is significant for {\it much smaller} values of the integration variables,
namely when $\phi^2\simeq {\sf C}(a)=M^{n-2}(Ma)^{2s(1-\xi)/\xi}$. In brief,
each integration variable sees two qualitatively different regions, one for
small values dominated by the auxiliary potential, and one for large values
controlled by the terms of the usual model. Of course, ``small'' and
``large'' are relative terms and may be influenced by a rescaling of the
variables of integration; for example, if we introduce
$\rho_k=(Ma)^s\phi_k$, then it follows that $\rho_k$ has an essential
support of order one or less. But in this case the influence of the
auxiliary potential occurs for extremely small values of $\rho_k$. As the
continuum limit is approached the disparate scales for the two effects
become even more pronounced. Incidentally, the importance of the very small
and very large integration domains for each variable would make a naive
application of either renormalization group arguments or Monte Carlo
calculations appear rather difficult.

In a formal continuum limit the density of the auxiliary potential term is
invariably of the form
  $$\frac{c\hbar^2}{\phi^2(x)}$$
for any value of $\xi$ and any space-time dimension $n$, where $c$ denotes a
suitable constant. We will present arguments that show that this term should
not be interpreted as a renormalization counterterm for the nonlinear
interaction term, but rather as a necessary renormalization for the {\it
kinetic energy term} with which it shares an identical scaling behavior.
(The analog of the centrifugal potential arising out of the kinetic energy
in nonrelativistic quantum mechanics is an appropriate one here.)

Such renormalization counterterms have been encountered in previous models
studied by the author, and one may expect that some of the general
discussion that held true for previous models may well hold true in the
present case \cite{k73}. In particular, we have in mind the implications of
such a counterterm for: (i) the presence of a hard-core interaction; (ii)
the limiting behavior as the nonlinear coupling constant is turned off being
not the usual free theory but rather a pseudo-free theory which retains the
essential effects of the hard-core potential; and (iii) the existence of a
meaningful perturbation theory in the nonlinear coupling constant, not about
the free theory, but instead about the pseudo-free theory.
\section{Lattice Regularized Models: First Formulation}
In this section we shall present the first of two formulations of a class of
models each of which we believe has the right behavior so that in the
continuum limit a non-Gaussian Euclidean quantum field theory emerges. Based
on the discussion in the preceding section it is possible that in Minkowski
space these models exhibit nontrivial scattering as well. Among such models
are natural candidates for nontrivial theories of $\varphi^4_n$ in
space-time dimensions $n\geq 5$ and perhaps $n=4$ as well. We begin this
section with some remarks on the difference between locality on the lattice
and locality in the continuum.
\subsubsection*{Lattice and continuum locality}
Let us illustrate the point of this subsection with a very simple example,
namely one that deals with Gaussian white noise. Suppose we want to find a
lattice prescription for the Gaussian functional integral given by
  $$e^{-\frac{1}{2}\!\int h^2(x)\,\dn}=\int e^{i\int
h(x)\phi(x)\,\dn}\,d\mu(\phi)\;.$$
The natural lattice formulation of this problem is
  $$e^{-\frac{1}{2}\Sigma_k h_k^2a^n}=N\int e^{i\Sigma_k
h_k\phi_ka^n}e^{-\frac{1}{2}\Sigma_k \phi_k^2a^n}\,\Pi d\phi_k\;,$$
in which the manifest locality---even ultralocality---of the continuum
result is already apparent in the lattice form. However, it is by no means
necessary that the lattice formulation needs to be local in order for the
continuum limit to be local. As an example we study the following
alternative lattice formulation for the same problem.

Let us consider
  $$N_\Xi\int e^{i\Sigma_k
h_k\phi_ka^n}e^{-\frac{1}{2}\Sigma_{k,l}\Xi_{k;l}\phi_k\phi_l a^n}\,\Pi
d\phi_k\;,$$
where $\Xi$ is a positive-definite matrix, is independent of the lattice
spacing, and has the property that $\Sigma_k\Xi_{k;l}=1$ for each fixed $l$.
 For general $\Xi$, the lattice exponent is certainly not local on the
lattice; however, the continuum limit leads to the same result as before.
Specifically,
  $$\lim_{a\ra 0}\:N_\Xi\int e^{i\Sigma_k
h_k\phi_ka^n}e^{-\frac{1}{2}\Sigma_{k,l}\Xi_{k;l}\phi_k\phi_l a^n}\,\Pi
d\phi_k=e^{-\frac{1}{2}\!\int h^2(x)\,\dn}\;.$$
An acceptable example for $\Xi$ is given by
  $$\Xi_{k;l}={\tilde M}\,\Pi_{j=1}^n\sin^2(k_j-l_j)/(k_j-l_j)^2\;,$$
where $\tilde M$ is a normalizing constant. This expression has the apparent
feature of being nonlocal. However, in the continuum any two points a finite
distance apart are actually an infinite number of (continuum) lattice steps
apart, and so there is no real nonlocality after all.
Insofar as the continuum limit is concerned, the matrix $\Xi_{k;l}$ is just
as good as the choice $\delta_{k,l}$. One matrix may be more natural
than the other, but it would be incorrect to say that one is ``right'' and
the other is ``wrong''. Since the continuum limit is the ultimate goal they
are both equally ``right''.

The purpose of this comment is to assuage any concern about continuum
nonlocality that may be raised in the lattice formulation to follow. What
appears to be nonlocal is in fact perfectly local in the continuum limit.
\subsection{Determining the Second Characteristic}
As a preliminary we choose a ``time'' direction in the Euclidean lattice;
this will become the time direction when one changes from a Euclidean to a
Minkowski theory. For this purpose we single out the first component $k_1$
of the lattice label as the future time direction, but for convenience we
shall simply refer to $k_1$ as the time direction. Evidently there are
$s=n-1$ space components, $k_2,\ldots,k_n$, remaining. We assume the
original lattice is a hypercube with an odd number, $2L+1$, of lattice sites
in each direction. Thus the total number of lattice points is
$N\equiv(2L+1)^{n}$, while the total number of lattice points in a spatial
slice (at constant $k_1$) is given by $N'\equiv(2L+1)^s$. We adopt periodic
boundary conditions in the spatial directions and Dirichlet boundary
conditions in the time direction. Let us next introduce a family of real
constants $\{\beta_k\}$ (with $k$ a lattice site label) each of which is
positive, $\beta_k>0$, and subject to the normalization condition
  $$\Sigma'_k\beta_k=1\;,$$  where the sum is over the $N'$ sites in any
spatial slice in the lattice, i.e., only $n-1$  dimensions of the lattice.
In doing the sum we hold $k_1$ fixed, and the normalization holds for any
$k_1$. We shall use a ``prime'' to signify when a sum, or a factor, refers
to a lattice of codimension one.  A suitable choice of constants $\beta_k$
is given in the form    $$\beta_k=\overline{\beta}_k\equiv
M'\,\Pi_{j=2}^nK^{-|k_j|}$$   where $K>1$ and $M'$ is a normalization factor
 given by
  $$M'=[(K-1)/(K+1-2K^{-L})]^{s}\;.$$

Armed with these definitions we introduce an ansatz for the weight function
appropriate to the second characteristic. In particular we let  $$
\psi_a(\phi)=J\frac{\,e^{-
w(\phi)}}{\Pi'_k[\Sigma'_l
\beta_{k-l}\phi_l^2]^{\gamma/2}}\;,$$
where $J$ is a normalization factor (to be chosen below), $w(\phi)$, with
$w(0)=0$, is a function designed to control the large field behavior, while
the factor in the denominator controls the small field behavior. We choose
the parameter $\gamma=1/2+\xi/(2N')$, $0<\xi<1$, in order, as we shall see,
that $$\int|\psi_a(\phi)|^2\,\Pi'd\phi_k=\infty\;,$$ while on the other
hand, along with a suitable $w$,
$$\int(\Sigma'g_k\phi_ka^s)^p|\psi_a(\phi)|^2\,\Pi'd\phi_k<\infty$$
 for all $p\geq1$. In terms of these expressions, the (negative of the)
second characteristic reads $$C_S(g)\equiv J^2\int[1-\cos(\Sigma'\phi_k
g_ka^s)]\frac{\,e^{-2w(\phi)}}{\Pi'_k[\Sigma'_l\beta_{k-l}\phi_l^2]^\gamma}
\,\Pi_k'd\phi_k\;.$$

Normalization is readily dispensed with. We choose a special ({\it sp}) test
sequence, for example we may choose
  $$g_k=g^{sp}_k\equiv M^{n/2}e^{-Ma\Sigma_{j=2}^n|k_j|}\;,$$
for a suitable constant $M$. This special sequence is chosen so that
$$\Sigma'(g^{sp}_k\,)^2a^s=M+O(a,L^{-1})$$
is finite and nonzero in the continuum limit. We declare that
$C_S(g^{sp})=1$, a condition that fixes the value of $J$, in particular, so
that $$J^{-2}\equiv\int[1-\cos(\Sigma'\phi_k
g^{sp}_ka^s)]\frac{\,e^{-2w(\phi)}}
{\Pi'_k[\Sigma'_l\beta_{k-l}\phi_l^2]^\gamma}\,\Pi_k'd\phi_k\;.$$

In order to better ascertain the continuum limit for the second
characteristic, it is convenient to introduce a new set of integration
variables which we shall refer to as {\it hyper-extreme spherical
coordinates}. Based on the definition $\phi_k\equiv\kappa\eta_k$,
$\kappa\geq0$, with $\Sigma'\phi_k^2\equiv\kappa^2$, let us consider the
measure
 \b \Pi'_k[\Sigma'_l\beta_{k-l}\phi_l^2]^{-\gamma}\,\Pi'_kd\phi_k\!\!\!\!\!\!
&&=\Pi'_k[\Sigma'_l\beta_{k-l}\phi_l^2]^{-\gamma}\,\Pi'_kd\phi_k\,
\delta(\kappa^2-\Sigma'_k\phi_k^2)\,d\kappa^2\\ &&=\kappa^{N'-2}\,
\kappa^{-N'-
\xi}\,d\kappa^2\,\delta(1-\Sigma'\eta_k^2)
\Pi'_k[\Sigma'_l\beta_{k-l}\eta_l^2]^{-\gamma}\,\Pi_k'd\eta_k\\
&&=2\kappa^{-1-\xi}\,d\kappa\,d\tau(\eta)\;,  \e
where $$d\tau(\eta)\equiv\delta(1-\Sigma'_k\eta_k^2)\,[\Pi'_k[\Sigma'_l
\beta_{k-l}\eta_l^2]^{-\gamma}\,\Pi_k'd\eta_k\;.$$
In terms of these variables it follows that  $$C_S(g)=2J^2\int[1-\cos(\kappa
\Sigma'\eta_kg_ka^s)]\,e^{-
2w(\kappa\eta)}
\kappa^{-1-\xi}\,d\kappa\,d\tau(\eta)\;,$$
and more particularly for the moments that
  \b <(\Sigma'\,
g_k\phi_ka^s)^p>&=&2J^2\int(\Sigma'\,g_k\eta_ka^s)^p\,\Pi'_k
[\Sigma'_l\beta_{k-l}\eta^2_l]^{-\gamma}\,e^{-2w(\kappa\eta)}\\
&&\;\;\;\;\:\:\;\times\,\kappa^{p-1-\xi}\,d\kappa\,\delta(1-\Sigma'
\eta^2_k)\,\Pi'_kd\eta_k\;. \e
Here we restrict attention to $0<\xi<1$ which is readily seen to lead to the
desired conditions that
$\int(\Sigma'\phi_kg_ka^s)^p|\psi_a(\phi)|^2\Pi'd\phi_k$ diverges for $p=0$
and is finite for $p\geq1$. The case where $\xi$ is larger is treated below.
\subsubsection*{First look at acceptable expressions for $w(\phi)$}
When moments exist, an acceptable expression for $w(\phi)$ is one for which
moments of the second characteristic are all of the same order of magnitude.
We can assure this by studying the quotient of any two moments and ensuring
that it neither vanishes nor diverges in the continuum limit. We focus our
initial attention on the quotient
$$\frac{<(\Sigma'g_k\phi_ka^s )^4>}{<(\Sigma'g_k
\phi_ka^s )^2>}=\frac{\int(\Sigma'\,g_k
\eta_ka^s)^4\kappa^{3-\xi}\,e^{-2w(\kappa\eta)}\,d\kappa\,d\tau(\eta)}
{\int(\Sigma'\,g_k\eta_ka^s
)^2\kappa^{1-\xi}\,e^{-2w(\kappa\eta)}\,d\kappa\,d\tau(\eta)}\;.$$   Suppose
that $w=\kappa^2/T$, then it follows that  \b\frac{<(\Sigma'g_k
\phi_ka^s)^4>}{<(\Sigma'g_k\phi_ka^s )^2>}&=&c\,T\,\frac{\int(\Sigma'\,g_k
\eta_ka^s)^4\,d\tau(\eta)}{\int(\Sigma'\,g_k\eta_ka^s )^2\,d\tau(\eta)}\\
&=&3c\,T\,a^{s}\frac{(\Sigma'_kg_k^2a^{s})(\Sigma'_lg_l^2a^{s})\int\eta_k^2
\eta_l^2\,d\tau(\eta)}{(\Sigma'_kg_k^2a^{s})\int\eta_k^2\,d\tau(\eta)}\;, \e
where $c$ denotes a fixed, nonzero, dimensionless constant. Now, as
discussed below, it is reasonable to assume something like
$$\frac{\int\eta_k^2\eta_l^2\,d\tau(\eta)}{\int\eta_m^2\,d\tau(\eta)}\approx
\beta_{k-l}\;.$$
As a consequence it is necessary to choose $T\propto (Ma)^{-2s }$---for some
fixed parameter $M$ with the dimension of a mass---in  order to have a ratio
that stays finite and away from zero as $a\ra\infty$ and $N'\ra\infty$. Note
that $T$ has the mass  dimensions of $\phi^2$, namely $n-2$. It is readily
seen that if $w=\kappa^4/T^2$, with $T\propto(Ma)^{-2s }$, a suitable
behavior is also found. In addition all other moment ratios are maintained
correctly with these same choices. Consequently, we conclude that the proper
choice for $w$ is given by
  $$w(\phi)=\Upsilon(\kappa\eta(Ma)^{s }\:)$$  for a large class of
functions $\Upsilon$ that may involve dimensional constants, but are
(effectively) {\it independent} of either the lattice spacing $a$ or the
number of lattice sites in a spatial slice $N'$. This remark has strong
consequences for the form of the regularized coefficients in the lattice
Hamiltonian and therefore in the lattice action.
\subsubsection*{Elementary exercises in many-variable integrations}
There are a number of significant differences in many-dimensional
integrations between Gaussian-like integrands and Poisson-like integrands of
the general kind that appear in the second characteristic. For this remark
we have in mind integrals of the form
 $$I_G(2p)=\int (\Sigma\phi_k^2)^p\,e^{-A\Sigma\phi_k^2}\,\Pi\,d\phi_k$$  as
well as those of the form  $$I_P(2p)=\int
(\Sigma\phi_k^2)^p\,e^{-A\Sigma\phi_k^2}[\Sigma\phi_k^2]^{-N/2}\,\Pi
\,d\phi_k\;,$$  where $p\geq 1$. Here, in this and the following exercise
section, we do not distinguish between $N$ and $N'$, and for simplicity in
notation we drop the prime. Note that the integrals in question are simple
``caricatures'' of the more complicated integrals of interest; nevertheless
these integrals will illustrate an important point. More relevant integrals
are studied below.

If we use the variables $\kappa$ and $\{\eta_k\}$ introduced previously, it
follows that
$$I_G(2p)=2\int\kappa^{2p}\,e^{-A\kappa^2}\kappa^{N-1}\,d\kappa\,\delta(1-
\Sigma\eta_k^2)\,\Pi\,d\eta_k\;.$$  Evaluation of this expression for large
$N$ is dominated by the factor $\kappa^{N-1}$, and a steepest descent method
can be used to evaluate to a suitable accuracy the integral over $\kappa$.
It follows that the stationary point for each integral of this form is given
to leading order by $\kappa=\sqrt{N/(2A)}$. As a consequence, for each
different value of $A$ the integrand lives on a different set---a disjoint
set of field values---as $N\ra\infty$ \cite{hid}. This fact is responsible
for divergences in perturbation calculations. For example, if we shift the
value of $A$ and attempt to calculate
$$I'_G(2)=2\int\kappa^{2}\,e^{-A'\kappa^2}\kappa^{N-1}\,d\kappa\,\delta(1-
\Sigma\eta_k^2)\Pi\,d\eta_k $$ in terms of $I_G(2p)$ by means of a
perturbation series in $\Delta A\equiv A'-A$, then we are led to the series
 $$I'_G(2)=I_G(2)-\Delta A I_G(4)+\textstyle{\frac{1}{2}}(\Delta A)^2
I_G(6)-\textstyle{\frac{1}{6}}(\Delta A)^3 I_G(8)+\cdots$$  which exhibits
{\it divergences} as $N\ra\infty$ since $I_G(2p)/I_G(2)\propto N^{(p-1)}$.
In the same spirit if one attempted to evaluate
$$I''_G=2\int\kappa^{2}\,e^{-A\kappa^2-B\kappa^4\Sigma\eta_k^4}\,
\kappa^{N-1}\,d\kappa\,\delta(1-\Sigma\eta_k^2)\,\Pi\,d\eta_k$$ by means of
a perturbation in $B$, then one would be led to the series
\b I''_G&=&I_G(2)-BI_G(6)\int(\Sigma\eta_k^4)\,d\sigma(\eta)\\&&+{\half}
B^2I_G(10)\int(\Sigma\eta_k^4)^2\,d\sigma(\eta)+\cdots\;,  \e
where we have introduced
$$d\sigma(\eta)\equiv\frac{\delta(1-\Sigma\eta_k^2)\,\Pi\,d\eta_k}{\int
\delta(1-\Sigma\eta_k^2)\,\Pi\,d\eta_k}\;.$$
Here again we encounter divergences, e.g., because $I_G(6)/I_G(2)\propto
N^2$ while $$\int(\Sigma\eta_k^4)\,d\sigma(\eta)\propto N^{-1}$$ and so the
net value for the first-order correction relative to the initial term is of
order $N$. In point of fact, the kind of divergences under discussion here
are extremely simplified examples---analogous to the ``caricatures'' spoken
about above---of the kind of divergences that arise in perturbative
calculations in relativistic $\varphi^4$ theory!

We now turn our attention to  a study of $I_P$. The expressions that we
shall study are once again caricatures of the more complicated ones studied
in previous sections; nevertheless they contain the essential ingredients
for present purposes. In particular, for $p\geq 1$, let us  examine
\b I_P(2p)&=&\int
(\Sigma\phi_k^{2})^p\,e^{-A\Sigma\phi_k^2}[\Sigma\phi_k^2]^{-N/2}\,\Pi\,
d\phi_k\\
&=&2\int\kappa^{2p}\,e^{-A\kappa^2}[\kappa^2]^{-N/2}\kappa^{N-1}\,d\kappa\,
\delta(1-\Sigma\eta_k^2)\,\Pi\,d\eta_k\\ &=&2\int\kappa^{2p-1}
e^{-A\kappa^2}\,d\kappa\,\delta(1-\Sigma\eta_k^2)\,\Pi\,d\eta_k \;. \e
Now we see no large power of $\kappa$ in the integrand to lead to
$N$-dependent factors. Indeed,
\b\frac{I_P(4)}{I_P(2)}&=&\frac{\int\kappa^{3}
e^{-A\kappa^2}\,d\kappa\,\delta(1-\Sigma\eta_k^2)\,\Pi\,d\eta_k}{\int\kappa
e^{-A\kappa^2}\,d\kappa\,\delta(1-\Sigma\eta_k^2)\,\Pi\,d\eta_k}\\
&=&\frac{\int\kappa^{3} e^{-A\kappa^2}\,d\kappa}{\int\kappa
e^{-A\kappa^2}\,d\kappa}=\frac{1}{A}\;, \e which is $N$-independent and
finite. With
 $$I'_P(2)=2\int\kappa
e^{-A'\kappa^2}\,d\kappa\,\delta(1-\Sigma\eta_k^2)\,\Pi\,d\eta_k$$
a perturbation expansion in $\Delta A$ as before is given by
$$I'_P(2)=I_P(2)-\Delta AI_P(4)+\half (\Delta A)^2
I_P(6)-\textstyle{\frac{1}{6}}(\Delta A)^3I_P(8)+\cdots$$ which exhibits no
divergences whatsoever. In a similar way
$$I''_P=2\int\kappa
e^{-A\kappa^2-B\kappa^4\Sigma\eta_k^4}\,d\kappa\,\delta(1-\Sigma\eta_k^2)\,
\Pi\,d\eta_k $$  admits an expansion in $B$ of the form
\b I''_P&=&I_P(2)-BI_P(6)\int(\Sigma\eta_k^4)\,d\sigma(\eta)\\ &&+{\half}
B^2I_P(10)\int(\Sigma\eta_k^4)^2\,d\sigma(\eta) +\cdots\;.  \e
 The $I_P$ coefficients are all finite in this case, but we still have
\b &&\int(\Sigma\eta_k^4)\,d\sigma(\eta)\propto N^{-1}\;,\\
&&\int(\Sigma\eta_k^4)^2\,d\sigma(\eta)\propto N^{-2}\;.  \e  Consequently,
if we reinterpret $B$ as being proportional to $N$, say $B={\o B}N$, then
each term of the perturbation series makes perfectly good sense as a power
series in $\o B$.

These simple examples serve to illustrate one aspect of the profound
difference that exists between many-variable integrations involving
Gaussian-like integrands compared with those that arise with Poisson-like
integrands. However, there is one glaring deficiency in our discussion,
namely we have not included any examples in which field derivatives play a
role. Let us remedy that situation.
\subsubsection*{Advanced exercises in many-variable integrations}
Let us again return to Gaussian-like integrals, this time some that resemble
scalar field models, namely
 \b{\cal I}_G&=&\int
e^{-A\Sigma(\phi_{k^*}-\phi_k)^2a^{(n-2)}-B\Sigma\phi_k^2a^n}\,\Pi
d\phi_k\\&=&2\int
e^{-A\kappa^2\Sigma(\eta_{k^*}-\eta_k)^2a^{(n-2)}-B\kappa^2a^n}\kappa^{N-1}
\,d\kappa\,\delta(1-\Sigma\eta_k^2)\,\Pi\, d\eta_k\;.  \e  The subsequent
calculation depends on the number of space-time dimensions involved. Let us
start simply and take $n=1$, namely
$${\cal I}_G=2\int
e^{-A\kappa^2\Sigma(\eta_{k+1}-\eta_k)^2a^{-1}-B\kappa^2a}\kappa^{N-1}\,d
\kappa\,\delta(1-\Sigma\eta_k^2)\,\Pi\, d\eta_k \;. $$   With $A\approx
B\approx 1$, it follows that $\kappa\approx\sqrt{N}$, $\eta_k^2\approx 1/N$,
and $(\eta_{k+1}-\eta_k)^2\approx a/N$. This is the case of Brownian motion
and it is interesting to observe that in the continuum limit the time
derivative of $\eta$ is almost everywhere finite. This behavior is unlike
that for $\phi$ for which
$$<(\phi_{k+1}-\phi_k)^2>=<\kappa^2(\eta_{k+1}-\eta_k)^2>\approx N(a/N)=
a\;.$$  It is only when the amplitude factor ($\kappa$) is included that the
paths pass from almost everywhere differentiable to nowhere differentiable!
Now let us turn our attention to higher dimensions, in particular $n\geq 3$
(ignoring logarithmic behavior characteristic of $n=2$). In that case---and
assuming for simplicity that $Na^n\approx 1$---it follows that
$\phi_k^2\approx a^{-(n-2)}$, $\kappa\approx N^{(1-1/n)}$, $\eta_k^2\approx
N^{-1}\approx a^n$, and  $(\eta_{k^*}-\eta_k)^2\approx a^n$ as well. It
follows, for example that $<(\phi_{k^*}-\phi_k)^2>\approx a^{-(n-2)}$, which
shows the divergence of the field difference comes about solely from the
radius factor $(\kappa)$ and is not at all due to the variables $(\eta_k)$.

If we add a quartic interaction term we are led to consider
$${\cal I}_G''=2\int
e^{-A\kappa^2\Sigma(\eta_{k^*}-\eta_k)^2a^{n-2}-B\kappa^2a^n-C\kappa^4\Sigma
\eta_k^4a^n}\kappa^{N-1}\,d\kappa\,\delta(1-\Sigma\eta_k^2)\,\Pi d\eta_k \;.
 $$  In this case the relative order of magnitude of the terms in the
principal part of the integrand is given by
$$e^{-A\kappa^2\Sigma(\eta_{k^*}-\eta_k)^2a^{n-2}-B\kappa^2a^n-C\kappa^4
\Sigma\eta_k^4a^n}\approx e^{-AN-BN^{(1-2/n)}-CN^{(2-4/n)}}\;,$$  which has
the interesting corollary that the quartic interaction term becomes
comparable with the kinetic term when $n=4$ and actually dominates the
kinetic term when $n\geq 5$. This is how (strictly) renormalizable and
nonrenormalizable models appear in this abbreviated language.

Now let us make a comparable study for Poisson-like integrals. Specifically,
and with $p\geq 1$, we consider  $${\cal I}_P(2p)=2\int(\Sigma
h_k\eta_ka^n)^{2p}\,e^{-A\kappa^2\Sigma(\eta_{k^*}-\eta_k)^2a^{n-2}-B
\kappa^2a^n}\kappa^{2p-1}\,d\kappa\,\delta(1-\Sigma\eta_k^2)\,\Pi
d\eta_k\;.$$  Our criterion of quality in the present case will relate to
the comparable
magnitude of the even moments. To study this issue we examine
$$\frac{{\cal I}_P(4)}{{\cal I}_P(2)}=\frac{\int(\Sigma
h_k\eta_ka^n)^4\,e^{-A\kappa^2\Sigma(\eta_{k^*}-\eta_k)^2a^{n-2}-B\kappa^2
a^n}\kappa^{3}\,d\kappa\,\delta(1-\Sigma\eta_k^2)\,\Pi d\eta_k}{\int(\Sigma
h_k\eta_ka^n)^2\,e^{-A\kappa^2\Sigma(\eta_{k^*}-\eta_k)^2a^{n-2}-B\kappa^2
a^n}\kappa\,d\kappa\,\delta(1-\Sigma\eta_k^2)\,\Pi d\eta_k}\;.$$
To approximate this quotient, we use the estimates that
$\eta_k^2\approx(\eta_{k^*}-\eta_k)^2\approx 1/N$, which for $n\geq3$ hold
just as before. Thus, in a self-evident notation, the quotient is
approximated by
\b \frac{{\cal I}_P(4)}{{\cal I}_P(2)}&\approx& 3a^{n}\frac{\int
( \Sigma_{k,l}h_k^2a^nh_l^2a^n)<\eta_k^2\eta_l^2>\int
e^{-A\kappa^2a^{n-2}-B\kappa^2a^n}\kappa^3\,d\kappa}{\int
(\Sigma_kh_k^2a^n)<\eta_k^2>\int
e^{-A\kappa^2a^{n-2}-B\kappa^2a^n}\kappa\,d\kappa}\\
&\propto&\frac{a^n}{N}\,\frac{\int
e^{-A\kappa^2a^{n-2}-B\kappa^2a^n}\kappa^3\,d\kappa}{\int
e^{-A\kappa^2a^{n-2}-B\kappa^2a^n}\kappa\,d\kappa} \;.
\e  For small $a$ and large $N$, the prefactor $a^n/N$ is tiny, and the only
way to counter this tiny factor is to allow significant weight to large
$\kappa$ values by making $A$ and $B$ small. To obtain the desired result
for the quotient, first imagine that $A=0$. In that case $B\propto 1/N$
leads to a quotient that is independent of $a$ and $N$. If instead $B=0$
then we may choose $A\propto a^2/N$ which again leads to a quotient that is
$a$- and $N$-independent. When both terms are present the same values prove
satisfactory. Observe that no reweighting of field support to disjoint sets
occurs if either parameter is rescaled by a finite amount. This fact is
reflected in the statement that an expansion in powers of $B$ or of $A$
would not lead to divergences order by order. (If this property were to
carry over to the type of integrals that arise for the models that are the
main subject of this paper, then it may well happen that for such models a
perturbation expansion in an interaction term does not lead to divergences
order by order.) For completeness, for the very special examples of this
section, we observe that if we had added another interaction term in the
exponent, such as $C\kappa^4\Sigma\eta_k^4a^n$, then the correct choice in
that case would be $C\propto a^n/N^2$. If instead the added term were
$D\kappa^{44}\Sigma\eta_k^{44}a^n$, then the correct choice would be
$D\propto a^{21n}/N^{22}\;!$

In summary, except for the term involving lattice derivatives, each of the
factors is correctly given by the observation that $\kappa$ enters in the
form $\kappa\sqrt{(Ma)^n/N}$. For the kinetic term, besides
$\kappa^2a^{(n-2)}/N$, an additional factor of $(Ma)^2$ is needed as well.
Insofar as an acceptable caricature of the lattice action suitable for
``large'' field values in the Poisson case goes, we conclude that we may
choose the integrand to be
$$e^{-\frac{1}{2}((Ma)^2/N)\Sigma(\phi_{k^*}-\phi_k)^2a^{(n-2)}-\frac{1}{2}
(m^2/N)\Sigma\phi_k^2a^n-g((Ma)^n/N^2)\Sigma\phi_k^4a^n}\;,$$  where $m^2$
and $g$ are cutoff-independent parameters.

\subsubsection*{Further analysis of acceptable expressions for $w(\phi)$}
We note that the examples just studied are not directly relevant for the
main problems of interest because we have, for clarity, omitted from
consideration in this simplified model the basic factor
$\Pi'_k[\Sigma'_l\beta_{k-l}\eta_l^2]^{-\gamma}$. It is straightforward to
see that this factor reweights the $\{\eta_k\}$ variables already
constrained to lie on the unit sphere in $N'$ dimensions in such a way that
points where $\eta_k^2=\delta_{kl}$, for each lattice site label $l$, are
significantly enhanced. That is, the measure $d\tau$ is heavily weighted on
the unit sphere along each coordinate axis, and significantly reduced away
from any coordinate axis. As a consequence, and in the absence of field
gradients, we have [with $\<(\cdot)\>\equiv\int(\cdot)d\tau(\eta)/\int
d\tau(\eta)$]
  \b\<\eta_k^2\>&\equiv&\frac{1}{N'}\;,\\
\<\eta_k^2\eta_l^2\>&\approx&\frac{\beta_{k-l}}{N'}+{\rm l.o.t.}\;,\\
\<\eta_k^2\eta_l^2\eta_m^2\>&\approx&\frac{\beta_{k-l}\beta_{l-m}\beta_{k-m}
}{N'}+{\rm l.o.t.}\;,  \e
etc., which crudely represent the fact that only very short range
correlations exist due to the strong peaking along coordinate axes, and the
only factors available to characterize that short range dependence are the
$\beta_k$ terms. The first equality is an identity, and the second one is
scaled to satisfy the first as a sum rule. The third equality fails to
satisfy a sum rule but is of the right order of magnitude. There are other
terms not explicitly written and denoted by l.o.t. (lower order terms) and
these involve potentially long range contributions but which are of a very
small magnitude, so small in fact that they will make {\it no} contribution
in the continuum limit. In the exercises treated above, for which the
$\beta$-reweighting factor was omitted, the weight of the $\eta$ variables
was distributed uniformly over the unit sphere, and that gave rise to
qualitatively different correlation functions than we are here discussing.
The present correlation functions have the feature that as the points
separate, i.e., $|k-l|\ra L$, the correlations effectively vanish, while in
the uniformly distributed case that was not the case. In the present case it
is actually {\it essential} that the correlations vanish so that our
ultimate field theory of interest can satisfy the cluster property (here in
spatial directions only).

Now let us consider the effect of introducing field derivatives. The effect
of field derivatives is to introduce additional, and longer-range
correlations into the direction field. Thus, as a consequence, we may expect
that the correlations discussed above are transformed into
  \b\<\eta_k^2\>&\equiv&\frac{1}{N'}\;,\\
\<\eta_k^2\eta_l^2\>&\approx&\frac{e^{-{\o M}a|k-l|}({\o M}a)^{s }}{N'}+
{\rm l.o.t.}\\
\<\eta_k^2\eta_l^2\eta_m^2\>&\approx&\frac{e^{-
{\o M}a(|k-l|+|l-m|+|k-m|)}({\o M}a)^{2s }}{N'}+{\rm l.o.t.}\;,  \e
for some mass parameter ${\o M}$.

We are now in a position to discuss the addition of field derivatives to the
factor $w(\phi)$ in the second characteristic weight function. Based on the
way that $w$ enters the lattice potential, it is clear that our original
identification of $w(\phi)=\Upsilon(\phi(Ma)^{s })$ remains a valid
conclusion when nonderivative terms are involved. The behavior of lattice
derivatives for space-time dimensions $n\geq 3$ is such that the derivative
of the direction field $\eta$ diverges, even without the need for large
$\kappa$ values; this conclusion stems from the fact that
$(\eta_{k^*}-\eta_k)^2\approx\eta_k^2$ itself.  Thus the classically
motivated factor $a^{(n-2)}$ is incorrect on the lattice (and more
particularly in the continuum limit) with respect to the factor $a^{-2}$.
For dimensions $n\geq 3$, this factor must be renormalized away by an
additional factor of $a^2$, just as was the case in the ``advanced
exercises''. For the nonderivative terms we are instructed to replace $\phi$
by $\phi(Ma)^{s }$; to augment this recipe in the presence of derivatives we
need only {\it drop} the classically motivated $a^{-2}$ factor in forming
the lattice derivative. Consequently, even in the presence of derivatives,
the renormalization of $\phi$ to become $\phi(Ma)^{s }$ is the only
modification required to obtain a model with cutoff-independent parameters!
With this prescription accepted then the inclusion of terms with derivatives
does not qualitatively change the discussion given earlier about how to
choose $w$.

The route to these conclusions has not been straightforward because the road
traveled is not a familiar one. Indeed, it is quite possible that we have
been too cavalier in our analysis of the needed renormalization for terms
involving the field gradient; the appropriate factors can no doubt be
determined by a more careful study. However, for the sake of further
discussion we shall adopt our present arguments.
\subsubsection*{Relaxation of sharp time requirement}
Up to this point we have limited the parameter $\xi$ so that $0<\xi<1$. By
our consistent use of ``cosine'' in the second characteristics, it is not
difficult to see that we could also allow $0<\xi<2$ with no real change.
Beyond this value we run into serious difficulty because the lowest order
even moment is no longer finite. Suppose, however, that we dealt not with
the characteristic function for the field but for a ``renormalized'' {\it
cube} of the field. By this expression we mean to consider
$$\exp\{-J^2\int[1-\cos(\Sigma'g_k\phi^3_ka^s)]\frac{e^{-2w(\phi)}\,\Pi'\,
d\phi_k}{\Pi'[\Sigma'\beta_{k-l}\phi^2_l]^\gamma}\}\;.$$
In this relation it is clear that the parameter $\xi$ that enters into
$\gamma$ in the usual way can be extended now so that $0<\xi<6$ and the
indicated expression will still be well defined. The larger values of $\xi$
do not support sharp-time fields, although they do support sharp-time fields
cubed in the manner shown. Even larger values of $\xi$ may be entertained by
considering still higher-order renormalized field powers. In brief,
therefore, the class of models can be extended beyond those that support
sharp-time fields merely by extending the range of the parameter $\xi$.

However, for convenience, we shall continue to restrict our analysis to the
case $0<\xi<1$ for which the smeared sharp-time field is an operator.
\subsubsection*{Continuum limit}
We now have accumulated enough information in order to take a continuum
limit for certain of the models in question. First we consider a
``primordial'' model in which $w(\phi)\equiv0$, and we consider the second
characteristic given in this case by the expression
$$C^o_S(g)\equiv2J^{o\,2}\int[1-\cos(\kappa\Sigma'\eta_k
g_ka^s)]\kappa^{-1-\xi}\,d\kappa\,d\tau(\eta)\;.$$
To make further progress we need to investigate the nature of the support of
the measure $\tau$. Recall that earlier we have asserted that the weighting
of the measure $\tau$ is concentrated along each coordinate axis on the unit
sphere. Therefore, it is not difficult to argue that in the continuum limit,
the measure $d\tau$ would be effectively the same as
  $$d\tau(\eta)\simeq d\tau_0(\eta)\equiv
F\,\Sigma'_l\,\delta(1-\eta^2_l)\,d\eta_l\,\Pi''_{k\neq
l}\,\delta(\eta_k)\,d\eta_k\;,$$
where $F$ denotes a suitable normalization. Use of this expression for the
second characteristic leads to
  \b C^o_S(g)\!\!\!\!\!\!&&=FJ^{o\,2}\Sigma'_l\int[1-\cos(\kappa
g_la^s)]\kappa^{-1-\xi}\,d\kappa\\
    &&=FJ^{o\,2}\Sigma'_l(|g_l|a^s)^\xi\,\int[1-\cos(v)]v^{-1-\xi}\,dv\\
   &&=\frac{\Sigma'_l|g_l|^\xi a^s}{\Sigma'_l|g^{sp}_l|^\xi a^s}\;, \e
which in the continuum limit becomes
  $$\lim_{a\ra0}C^o_S(g)=K^o\int|g({\bf x})|^\xi\,d^s\!x\;,$$
where $(K^o)^{-1}\equiv\int|g^{sp}({\bf x})|^\xi\,d^s\!x$.

Next, as a further example, we include one particular term in $w$ and again
take the continuum limit. In particular, we examine the case where
$$C_S(g)=2J^2\int[1-\cos(\kappa\Sigma'\eta_kg_ka^s)]e^{-(Ma)^{2s}\kappa^2}\,
\kappa^{-1-\xi}\,d\kappa\,d\tau(\eta)\;.$$
The same discussion regarding the general behavior of the measure $\tau$
holds again, and we conclude that
  $$C_S(g)=FJ^2\Sigma'_l\int[1-\cos(\kappa
g_la^s)]e^{-(Ma)^{2s}\kappa^2}\,\kappa^{-1-\xi}\,d\kappa\;,$$
which implies for the continuum limit that
  $$\lim_{a\ra0}C_S(g)=K\int d^s\!x\int\{1-\cos[\varrho g(
{\bf x})]\}e^{-M^{2s}\varrho^2}\,\varrho^{-1-\xi}\,d\varrho\;,$$
where
  $$K^{-1}\equiv\int d^s\!x\int\{1-\cos[\varrho g^{sp}(
{\bf x})]\}e^{-M^{2s}\varrho^2}\,\varrho^{-1-\xi}\,d\varrho\;.$$
\section{Lattice Regularized Models: Second Formulation}
In the preceding section we have made an ansatz for the weight function of
the second characteristic by choosing the form of $\psi_a(\phi)$. We also
made considerable progress in characterizing acceptable functions
$\psi_a(\phi)$, as preliminaries to finding the lattice ground state
$\Psi_a(\phi)$ which are related by formulas presented in Section 1.
Unfortunately, it is all but impossible to find analytic expressions for
$\Psi_a$ from the given form of $\psi_a$. Thus to proceed further we must
find an alternative way to analytically express the lattice ground state
$\Psi_a$ so that we can define the lattice Hamiltonian and the lattice
action. We will do so by choosing a different starting point, but one which
has the {\it same continuum limit} for the second characteristic as just
discussed.

In our second formulation we do not start with the weight function for the
{\it second characteristic} of the sharp-time characteristic function but
rather with the weight function for the {\it characteristic function}
itself. In other words, rather than initially assume a form for $\psi_a$ we
shall make an ansatz regarding $\Psi_a$ directly. We are unable (and not
necessarily interested) to determine whether or not the distribution of the
sharp-time field is a generalized Poisson distribution on the lattice (as
was the case previously), but since the continuum limit will agree for the
two different starting points it follows that the continum limit in the
second case is indeed described by a generalized Poisson distribution.

In the present case we postulate the form of the lattice ground state to be
  $$\Psi_a(\phi)=\frac{e^{-W(\phi)}}{\Pi'_k[\phi_k^2+
{\sf C}(a)]^{(1+\xi)/4}}\;,\;\;\;\;\;\;\;\;{\sf C}(a)>0\;,$$
where $W$ includes the necessary constant to ensure that
$\int|\Psi_a(\phi)|^2\Pi'_k\,d\phi_k=1$. This starting point may look fairly
standard accept for the presence of the product factor in the denominator.
Observe in the present case there is no appearance of the $\beta$
parameters. Our goal now is to show that this choice leads to the same
continuum limit in the two special cases discussed previously.

First we assume that $W(\phi)\equiv c$, $c$ a constant, a case that
corresponds to our ``primordial'' situation. In that case we consider the
characteristic function given by
  $$C^o(g)={\sf N}^o\int e^{i\Sigma'
g_k\phi_ka^s}\,\Pi'_k\frac{\,d\phi_k}{[\phi_k^2+
{\sf C}(a)]^{(1+\xi)/2}}\;,$$
where ${\sf N}^o$, which arises from $c$, ensures normalization. By a change
of integration variables it follows that
  \b C^o(g)&&\!\!\!\!\!={\prod_k}' K^o_C\int e^{i\varrho
g_k}\,\frac{\,d\varrho}{[\varrho^2+a^{2s}{\sf C}(a)]^{(1+\xi)/2}}\\
   &&={\prod_k}'\{1-K^o_C\int[1-\cos(\varrho
g_k)]\,\frac{\,d\varrho}{[\varrho^2+a^{2s}{\sf C}(a)]^{(1+\xi)/2}}\}\;.  \e
Observe that
  $$(K^o_C)^{-1}=\int [\varrho^2+a^{2s}
{\sf C}(a)]^{-(1+\xi)/2}\,d\varrho=a^{-s\xi}{\sf C}(a)^{-\xi/2}F\;,$$
where $F\equiv\int(u^2+1)^{-(1+\xi)/2}\,du$ is a fixed number depending only
on $\xi>0$. We next choose the $a$ dependence of ${\sf C}(a)$ so that
$K^o_C\equiv 2J^2a^s$ for some constant $J$. This leads to
  $${\sf C}(a)=M^{n-2}(Ma)^{2s(1-\xi)/\xi}\;,$$
for some mass $M$. With this choice for $\sf C$ it follows that
  $$C^o(g)={\prod_k}'\{1-2J^{o\,2}a^s\int[1-\cos(\varrho
g_k)]\,\frac{\,d\varrho}{[\varrho^2+a^{2s}{\sf C}(a)]^{(1+\xi)/2}}\}\;.$$
Although ${\sf C}(a)$ may, depending on the value of $\xi$, converge to
zero, remain constant, or diverge to infinity as $a\ra0$, the combination
$a^{2s}{\sf C}(a)\propto a^{2s/\xi}$ always goes to zero as $a\ra0$.
Therefore, we finally conclude that
  \b \lim_{a\ra0}C^o(g)\!\!\!\!\!&&=\exp(\!\!(-2J^{o\,2}\int\{1-\cos[\varrho
g({\bf x})]\}|\varrho|^{-1-\xi}\,d\varrho)\!\!)\\
   &&=\exp[-K^o\int|g({\bf x})|^\xi\,d^s\!x]\;.  \e   If we again assert
that $C^o(g^{sp})=\exp(-1)$, then it follows that {\it the continuum limit
is exactly the same as was previously obtained}. Hence the ansatz for the
second form of regularization has been verified for the choice $W=$
constant.

 We next study the case $W=(Ma)^{2s}\Sigma'\phi_k^2+c$, where $c$, which is
different than before, is chosen to ensure normalization. The expression to
be studied reads
  \b C(g)\!\!\!\!\!&&={\sf N}\int
e^{i\Sigma'\phi_kg_ka^s}e^{-(Ma)^{2s}\Sigma'\phi_k^2}\,\Pi'\frac{\,d\phi_k}
{[\phi_k^2+{\sf C}(a)]^{(1+\xi)/2}}\\
  &&={\prod_k}' K_C\int e^{i\varrho
g_k}e^{-M^{2s}\varrho^2}\,\frac{\,d\varrho}{[\varrho^2+a^{2s}
{\sf C}(a)]^{(1+\xi)/2}}\\
  &&={\prod_k}'\{1-K_C\int[1-\cos(\varrho
g_k)]e^{-M^{2s}\varrho^2}\,\frac{\,d\varrho}{[\varrho^2+a^{2s}
{\sf C}(a)]^{(1+\xi)/2}}\}\;.  \e
In the present case
 \b (K_C)^{-1}\!\!\!\!\!&&=\int
e^{-M^{2s}\varrho^2}\,\frac{\,d\varrho}{[\varrho^2+a^{2s}
{\sf C}(a)]^{(1+\xi)/2}}\\
  &&=a^{-s\xi}{\sf C}(a)^{-\xi/2}\int e^{-M^{2s}[a^{2s}
{\sf C}(a)]u^2}\,\frac{\,du}{(u^2+1)^{(1+\xi)/2}}\;, \e   and in the
continuum limit the exponent in the integrand of the last integral makes no
contribution and thus asymptotically $K_C$ is exactly what it was in the
case when $W=$ constant!
Hence we may determine that the continuum limit in the present case is given
by
$$\lim_{a\ra0}C(g)=\exp(\!\!(-2J^2\int\{1-\cos[\varrho
g({\bf x})]\}\,e^{-M^s\varrho^2}\frac{\,d\varrho}
{|\varrho|^{1+\xi}})\!\!)\;,$$
and once again the value of $J$ is fixed by appealing to the special test
sequence $g^{sp}$. As a consequence, the continuum limit in the present case
cooincides with the second example studied at the end of Section 2. Observe
additionally that this equality of the continuum limit has been acheived
with {\it exactly the same renormalization of the additional term}---namely
$\exp[-(Ma)^{2s}\Sigma'\phi_k^2]$---as was used in the previous section.

Thus, based on the identical form of the continuum limits and the appearance
of an identical renormalization, we are persuaded in favor of adopting this
second and alternative regularization based on the choice of the lattice
ground state $\Psi_a(\phi)$ rather than choosing the weight function
$\psi_a(\phi)$ that entered the second characteristic in the first approach.
Although we have not studied any further choices of weight functions $w$ and
$W$ in the two cases, we strongly believe that the second regularization
captures the essence of what is contained in the first regularization.
Hereafter, we choose the second form of regularization exclusively.
\subsection{Determination of the Auxiliary Potential}
Having agreed that the ansatz
  $$\Psi_a(\phi)=\frac{e^{-W(\phi)}}{\Pi'_k[\phi_k^2+
{\sf C}(a)]^{(1+\xi)/4}}$$
provides an acceptable choice for a lattice ground state, we are in position
to determine the form of the auxiliary potential $P$. To focus wholly on the
auxiliary potential let us return to the ``primordial'' model for which
$W(\phi)=c$, namely for a ground state given by
  $$\Psi^o_a(\phi)=\frac{{\sf N}^{o\,1/2}}{\Pi'_k[\phi_k^2+
{\sf C}(a)]^{(1+\xi)/4}}\;.$$
With this form of ground state the auxiliary potential $P$ may be identified
by way of the reduced Hamiltonian
 \b {\cal H}^o_{a}&\equiv& -{\half}{\hbar^2}Y(a)^{-1}a^{-
s}[\Sigma'\frac{\partial^2}{\partial\phi_k^2}-
\frac{1}{\Psi^o_a(\phi)}\Sigma'\frac{\partial^2\Psi^o_a(\phi)}{\partial
\phi_k^2}]\\
&\equiv&-{\half}{\hbar^2}Y(a)^{-1}a^{-s}\Sigma'\frac{\partial^2}{\partial
\phi_k^2}\,+\,\Sigma' P(\phi_k,a)a^{s}\;.  \e    The result is that
  \b
P(\phi_k,a)&\equiv&{\half}{\hbar^2}Y(a)^{-1}a^{-2s}\,\frac{1}{\Psi^o_a(\phi)
}\frac{\partial^2\Psi^o_a(\phi)}{\partial\phi_k^2}\\
&=&{\half}{\hbar^2}Y(a)^{-1}a^{-2s}\,\frac{4^{-1}(1+\xi)(3+\xi)\phi_k^2-2^{-
1}(1+\xi){\sf C}(a)}{[\phi_k^2+{\sf C}(a)]^2}\\
&\equiv&\frac{{\sf A}(a)\phi_k^2-{\sf B}(a)}{[\phi_k^2+{\sf C}(a)]^2}\;, \e
 which is just the expression previously quoted in Section 1.2. Although we
have normally assumed units such that $\hbar=1$, in the preceding formula we
have incorporated the correct $\hbar^2$ factor explicitly in order to
establish---for this one time only---its proper role.

We note in passing that the ground state expectation of $P$ diverges in the
continuum limit when it formally becomes proportional to $\phi^{-2}(
{\bf x})$. Therefore the ground state expectation of the kinetic energy term
must also diverge so that the sum is zero in the present case. This
compensation of divergences demonstrates that the auxiliary potential serves
as a renormalization counterterm for the kinetic energy. This same
interpretation remains true even with the addition of other potential terms.
\subsection{Lattice Model Proposal}
It is now a small step to propose a lattice model for the $\varphi^4_n$
theory based on the auxiliary potential just identified. Let us first recall
the discussion related to the toy model in Section 1.1. There we learned
that the model without moments (based on the Cauchy distribution) led to a
one-dimensional potential analogous to the auxiliary potential in the
lattice model as identified with the help of the ``primordial'' model ($W=$
constant). We also recall that for the toy model involving the modified
Bessel function the weight function for the characteristic function has, for
very small $\rho$ values, a functional form for small $x$ identical to the
Cauchy distribution. This fact suggests that we define the model
by a lattice action which for small field values is equivalent to that of
the auxiliary potential and for large field values has the functional form
of the conventional action up to suitable bare parameters. In particular, we
choose the generating functional for the lattice Euclidean-space
$\varphi^4_n$ given by
  \b
  S(h)&=&N_0\int\exp(\!\!(\Sigma\, h_k\phi_ka^n-\half
Y(a)\Sigma(\phi_{k^*}-\phi_k)^2a^{(n-2)}-\half m_0^2(a)\Sigma\phi_k^2a^n\\
& &\;\;\;\;\;\;\;\;\;\;-g_0(a)\Sigma\phi_k^4a^n-\Sigma\,
  \{[{\sf A}(a)\phi_k^2-{\sf B}(a)]/[\phi_k^2+{\sf C}(a)]^2\}a^n)\!\!)\:\Pi
d\phi_k\,,  \e
where $\sf A$, $\sf B$, and $\sf C$ are given above. Observe that we have
not chosen to define the model by picking $W$ for the simple reason that we
do not have any analytic proposal for that quantity; rather we have returned
to the lattice action which preserves its simple analytic form even in the
presense of the auxiliary potential. Observe also that for small $Ma$ the
coefficients of the conventional action terms ($Y,m_0^2,$ and $g_0$) are all
small and this fact makes for a greater isolation of the large and small
field behavior. Analogously to the case of small $\rho$ in the second toy
model, this isolation of the two regimes gives even greater credence to our
choice of the auxiliary potential determined from the ``primordial'' model
as the form adopted when the conventional terms are present.

It may be natural for the reader to ask why we have not started with the
second formulation of Section 3 and dispensed with the first formulation of
the model in Section 2 all together. The reason we have included the first
formulation is because experience has shown that it is much easier to
motivate unusual forms of the weight function in the second characteristic
than in the characteristic function itself. Only after gaining familiarity
with the strange denominators in the weight function of the second
characteristic---{\it factors which are mandated in order to satisfy the
physically dictated moment requirements for the weight function of the
second characteristic}---and seen the consequences of these denominator
terms in the continuum limit, only then do the unusual denominators of the
weight function for the characteristic function of the sharp-time field
itself become reasonable and therefore acceptable.
\subsubsection*{A remark on O$(\o N)$ symmetric models}
It is interesting to observe that if one deals with an ${\o N}$-component
scalar field with an O$({\o N})$ symmetry, then the principal change
involved is the replacement of all factors $\phi_k^2$ by the O$(
{\o N})$-symmetric term $\Sigma_\alpha(\phi_k^\alpha)^2$,
$1\leq\alpha\leq{\o N}$, etc., along with a corresponding change of $\gamma$
or equivalently $\xi$. How this additional symmetry fits into the present
scheme may be partially traced from previous work \cite{zhu}.

\section*{Acknowledgments}
It is a pleasure to thank Klaus Baumann and Detlev Buchholz for their
comments on a preliminary version of this paper. The Doppler Institute,
Czech Technical University, Prague, is thanked for their hospitality during
which time a second version was largely completed.

\end{document}